\documentclass[twocolumn,aps,prl,showpacs,amsmath,superscriptaddress,longbibliography,notitlepage]{revtex4-2}
\usepackage{amssymb,mathtools}
\usepackage{mathrsfs}
\usepackage{physics}
\usepackage{graphicx}
\usepackage{float}
\usepackage[caption=false]{subfig}
\usepackage[pdftex,colorlinks=red]{hyperref}
\hypersetup{
	colorlinks,
	citecolor=magenta,
	linkcolor=magenta,
	urlcolor=magenta}
\usepackage{appendix}

\usepackage{epstopdf}

\usepackage[normalem]{ulem}
\usepackage{verbatim}
\usepackage{xcolor}
\usepackage{bm}

\usepackage{soul}
\usepackage{multirow}

\begin{document}

\title{Anomalous fractal scaling in two-dimensional electric networks}

\author{Xiao Zhang} 
\affiliation{School of Physics, Sun Yat-sen University, Guangzhou 510275, China}
\author{Boxue Zhang} 
\affiliation{School of Physics, Sun Yat-sen University, Guangzhou 510275, China}
\author{Haydar Sahin} \email{sahinhaydar@u.nus.edu}
\affiliation{Department of Electrical and Computer Engineering, National University of Singapore, Singapore 117551, Republic of Singapore}
\affiliation{Institute of High Performance Computing, A*STAR, Singapore 138632, Republic of Singapore}
\author{Zhuo Bin Siu}
\affiliation{Department of Electrical and Computer Engineering, National University of Singapore, Singapore 117551, Republic of Singapore}
\author{S M Rafi-Ul-Islam}
\affiliation{Department of Electrical and Computer Engineering, National University of Singapore, Singapore 117551, Republic of Singapore}
\author{Jian Feng Kong}
\affiliation{Institute of High Performance Computing, A*STAR, Singapore 138632, Republic of Singapore}
\author{Bing Shen} 
\affiliation{School of Physics, Sun Yat-sen University, Guangzhou 510275, China}
\author{Mansoor B. A. Jalil}
\affiliation{Department of Electrical and Computer Engineering, National University of Singapore, Singapore 117551, Republic of Singapore}
\author{Ronny Thomale}
\affiliation{Institute for Theoretical Physics, University of W\"urzburg, W\"urzburg D-97074, Germany}
\author{Ching Hua Lee}  \email{phylch@nus.edu.sg}
\affiliation{Department of Physics, National University of Singapore, Singapore 117542, Republic of Singapore}

\begin{abstract}
\noindent 
Much of the qualitative nature of physical systems can be predicted from the way it scales with system size. Contrary to the continuum expectation, we observe a profound deviation from logarithmic scaling in the impedance of a two-dimensional $LC$ circuit network. We find this anomalous impedance contribution to sensitively depend on the number of nodes $N$ in a curious erratic manner, and experimentally demonstrate its robustness against perturbations from the contact and parasitic impedance of individual components. This impedance anomaly is traced back to a generalized resonance condition reminiscent of the Harper's equation for electronic lattice transport in a magnetic field, even though our circuit network does not involve magnetic translation symmetry. It exhibits an emergent fractal parametric structure of anomalous impedance peaks for different $N$ that cannot be reconciled with continuum theory and does not correspond to regular waveguide resonant behavior. This anomalous fractal scaling extends to the transport properties of generic systems described by a network Laplacian whenever a resonance frequency scale is simultaneously present.
\end{abstract}
\date{\today}

\maketitle

\section{Introduction}

From dimensional analysis to the universality of critical phase transitions, scaling theory provides a universal paradigm for the principal understanding of most physical phenomena~\cite{fisher_theory_1967, stanley_scaling_1999, hilfer_scaling_1992, cardy_scaling_1996, frohlich_scaling_1983, chen_scaling_2016, zuo_scaling_2021}. Particularly interesting are ``marginal'' scenarios, where observables exhibit great freedom in their functional dependency on the physical variables~\cite{leigh1995exactly, muller_marginal_2015, dresselhaus_numerical_2021, zirnbauer_marginal_2021}. 
A classic example is the electrical impedance $Z$ of a $D$-dimensional sample of characteristic length $N$, which scales as $Z\sim N^{2-D}$; in particular, for $D=2$, $Z$ must scale slower than any power of $N$, most commonly logarithmically.

Indeed, logarithmic scaling is ubiquitous in physics, appearing in a broad range of contexts as disparate as conformal field theory, disorder Green's functions, strongly coupled quantum fields, and graph complexity~\cite{mackinnon1983scaling, poland_conformal_2019, zamolodchikov_exact_1989, lee_disordered_1985, albert_statistical_2002}. It represents the paradigmatic slower-than-power-law behavior that appears naturally in various scale-free scenarios. Particularly, impedance scaling in electrical circuits, as a function of circuit size, dutifully displays this scaling behavior when the circuit is either entirely reactive or resistive. For instance, the dimensionality of lattice models determines whether the impedance scales linearly, logarithmically or saturates to a constant impedance value. While 1D and 2D samples exhibit these scaling characteristics such as linear or logarithmic scaling,  the impedance in circuits characterized with dimensionality $D\geq3$ experiences rapid saturation proportional to $D$. Nevertheless, although the dimensionality of the lattice delineates the scaling characteristics, it ceases to be the predominant determinant in heterogeneous resonant medias. In fact, the scaling profile of $LC$ circuits with inductance $L$ and capacitance $C$ relies more on the form of the lattice array rather than the lattice dimension~\cite{chen_equivalent_2021,tan_electrical_2019}. This is because the impedance across two opposite farthest sites varies due to the parameter space irrespective of the lattice dimension. To explain this, we will conceptually elucidate and experimentally demonstrate the resonant conditions through a seemingly elementary physical 2D system by evaluating its corner-to-corner impedance behavior.
\begin{figure}
	\includegraphics[width=\linewidth]{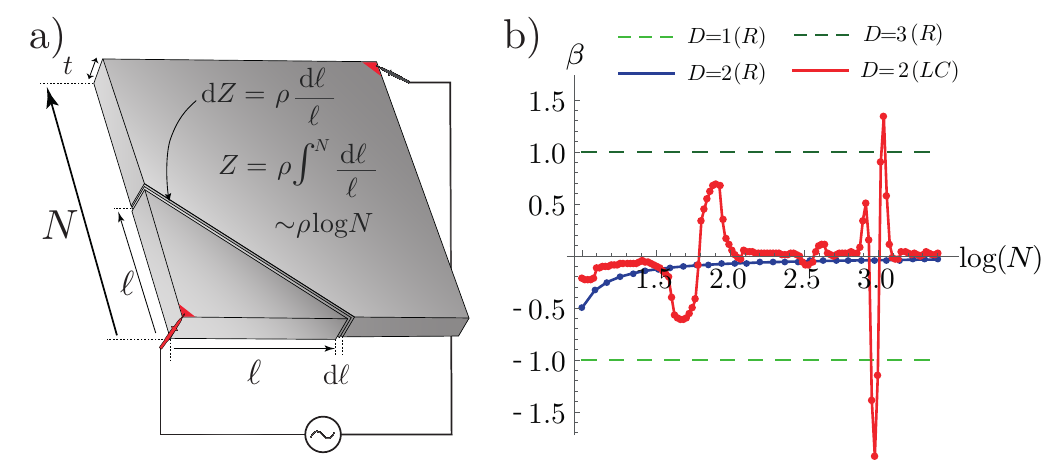}
	\caption{\textbf{Origin of logarithmic impedance scaling in continuum media and its violation in lattices.} (a) In a continuous sample such as a square plate of length $N$ and resistivity $\rho$, the diagonal-to-diagonal impedance necessarily scales like $\rho\log N$. This is easily seen by slicing the sample into strips perpendicular to the diagonal and noticing that each strip approximately contributes a serial impedance that is inversely proportional to its width. This is because each successive ``shell'' in the sample scales with its linear dimension $l$ as $l^{D-1}=l$, such that the total impedance scales like $\int^N l^{-1}dl\sim \log N$. 
	(b) Behavior of $\beta=-\frac{d\log |Z|}{d\log N}$, the fractional rate of change of impedance $Z$ diagonally with the system size $N$, across circuit lattices with and without a AC frequency scale. While a purely resistive 2D circuit (blue) exhibits a smoothly vanishing $\beta$ consistent with the continuum approximation in (a), our 2D $LC$ circuit (red) with an illustrative frequency scale of $\omega_r=1.95$ exhibits anomalous scaling behavior with pronounced and erratically located peaks. The dashed lines represent the constant or saturated scaling of other dimensions derived from $\beta$ of scaling theory applicable to non-resonant reactive media.
	}
\label{fig1}
\end{figure}

The two-point impedance and resistance problem has garnered significant attention~\cite{bartis_lets_1967,kirkpatrick_percolation_1973,venezian_resistance_1994,lavatelli_resistive_1972,zemanian_classical_1984,aitchison_resistance_1964,montroll_random_1965,morita_useful_1971} as it not only allows for the study of electrical conductivity, but also serves as a means of uncovering new physical phenomena related to lattice dimension, network model, lattice uniformity, and boundary design from the changes in the electrical characteristics in the presence of perturbations or disorder~\cite{asad_perturbed_2014,owaidat_perturbation_2016,giordano_disordered_2005,cserti_perturbation_2002,lee2021many}. The extensive research conducted in this expansive field has enhanced our fundamental understanding of electric circuits~\cite{izmailian_generalised_2014,tan_resistance_2017,cserti_uniform_2011,owaidat_interstitial_2010,asad_infinite_2005,doyle_random_2000,mamode_calculation_2019,pan_electric-circuit_2021,chen_electrical_2019,chen_electrical_2020,chen_electrical_2020-1,chen_electrical_2020-2,chen_electrical_2020-3,ammar_electrical_2022,tan_resistance_2017} and has practical applications in the design of various circuit systems, including topolectrical circuits~\cite{lee_topolectrical_2018,li2019emergence,wang_circuit_2020,wang2022observation,shang2022experimental,wu2023evidencing,zhang2023electrical}, non-linear systems~\cite{hohmann2023observation,tuloup2020nonlinearity,kotwal_active_2021,kengne_ginzburglandau_2022}, condensed matter counterparts~\cite{ningyuan2015time,rafi-ul-islam_topoelectrical_2020}, and metamaterials~\cite{kapitanova_photonic_2014}. In addition to numerical approaches such as the Laplacian formalism~\cite{wu_theory_2004,tzeng_theory_2006,cernanova_nonsymmetric_2014}, various analytical methods have been developed for determining the two-point impedance, including the recursion-transform method~\cite{tan_electrical_2019,tan_two-point_2016,tan_characteristic_2017,fang_circuit_2022,zhou_fractional-order_2017,tan_recursion-transform_2015,tan_recursion-transform_2015-2}, the lattice Green's function~\cite{joyce_exact_2017,cserti_application_2000,katsura_lattice_1971,mamode_revisiting_2021,joyce_exact_2002,joyce_1973}, asymptotic expansion~\cite{essam_exact_2009,izmailian_asymptotic_2010}, and the method of images~\cite{mamode_electrical_2017,sahin2022impedance}. While each of these methods employs a distinct approach to evaluate the impedance in both reactive and resonant media, they all require the circuit network to possess symmetries such as inversion and translation symmetries. The role of these symmetries has not been thoroughly explored in the literature, but their presence may result in anomalous behaviors that can be uncovered through the impedance scaling in electric circuits.

In this work, we report a dramatic uniform scaling violation in the impedance across $LC$ circuit lattices, resulting from the suppression of current at the boundaries due to the circuit symmetries. To demonstrate this, we specifically examine a 2D square $LC$ circuit, wherein its reactive counterpart displays a notable logarithmic scaling. Although the same qualitative picture can be observed in circuits with different dimensions, the 2D $LC$ circuit allows us to investigate the origin of uniform scaling violations using a simpler yet richer example. Naively, one would expect from the impedance of a 2D circuit to vary smoothly with the number of unit cells from its continuum analogue since the circuit lattice can be construed as a discretization of 2D conducting plates. However, while this indeed holds for non-resonant circuits, such as those containing capacitors or resistors exclusively, the impedance behavior for resonant media i.e., $LC$ circuits, cannot be more different. Our theoretical and experimental investigations reveal curious impedance enhancements of up to a few orders at certain lattice sizes $N$, whose roots can be traced to a new commensurability criterion associated with a Hofstadter butterfly-like fractal structure. This challenges the applicability of a continuum description in even the simplest of resonant media.

\section{Results} 

\begin{figure*}[]
	\centering
	\includegraphics[width=\linewidth]{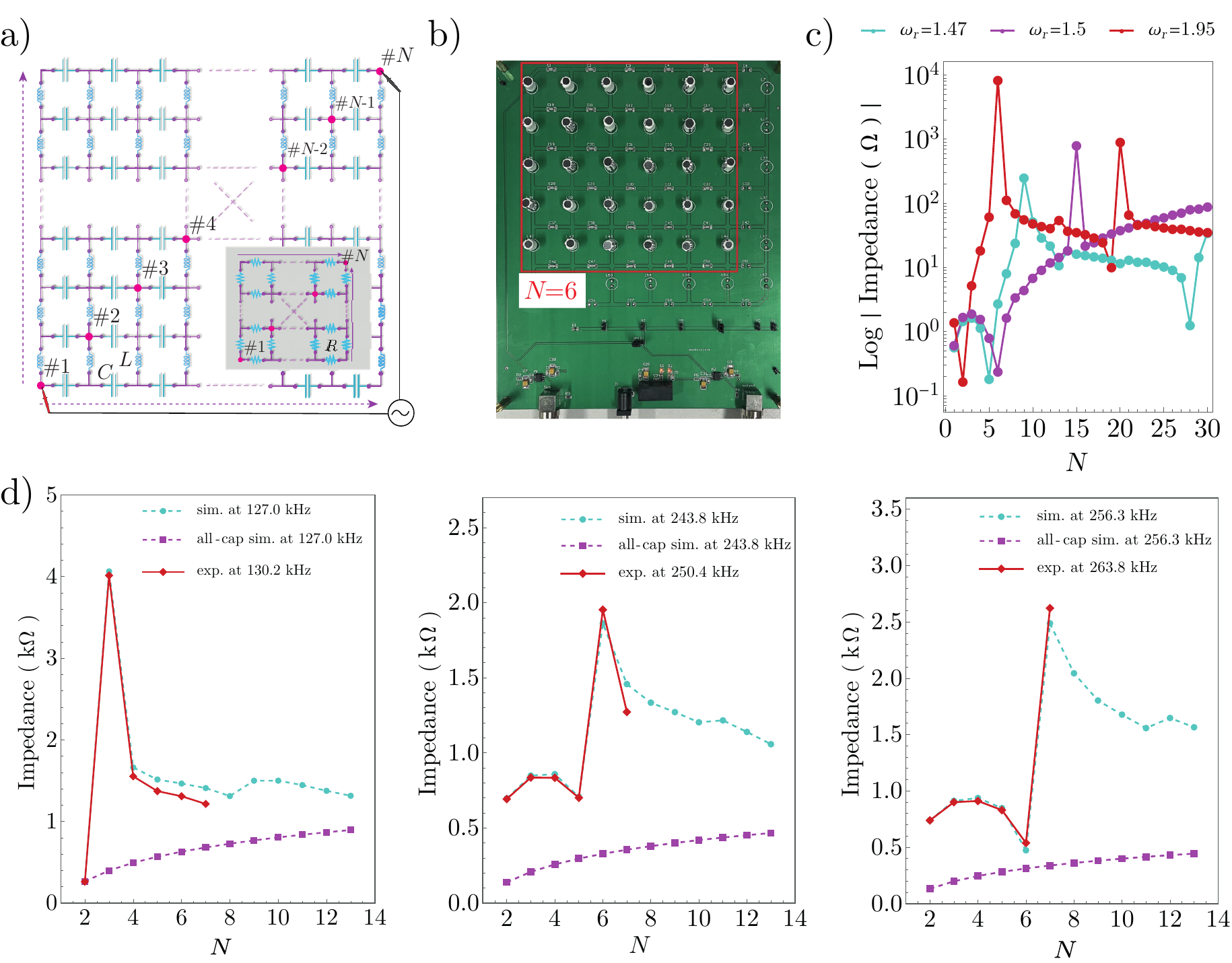}
	\caption{ \textbf{Circuit description and measured anomalous impedance scaling.} (a) Our circuit is a $N\times N$ square lattice array with the horizontal and vertical links being capacitors $C$ and inductors $L$ respectively. The corner-to-corner impedance $Z$ is measured by running a current between the lower left and upper right ($N$-th) nodes. The scaling behavior of $Z$ is revealed to contrast strongly with the logarithmic scaling of a similar but uniform circuit array consisting of only one type of element i.e., resistors (shown in the inset), or capacitors. 
	(b) We implement our $LC$ circuit arrays on circuit boards, and control the lattice size $N$ through switches. Shown here is the $6\times 6$ case - our board shown here admits up to the $N=7$ case. (c) In principle, with purely capacitive or inductive $LC$ components, the corner-to-corner impedance of our circuit becomes drastically higher by a few orders at particular lattice sizes $N$, and depends sensitively on $\omega_r=\omega\sqrt{LC}$ according to Eq.~\eqref{Z}. (d) These anomalous corner-to-corner impedance peaks are attenuated in our experimental measurements but are still robustly prominent, as shown in these plots at three illustrative AC frequencies $\omega$. The measured (exp, red) data agrees well with the simulated values (sim, cyan) with estimated parasitic resistances (estimated to be R$_{pL}$ = 3.3 $\Omega$, R$_{pC}$ = 4.5 $\Omega$, R$_{pW}$ = 0.1 $\Omega$, see Methods: Analysis of uncertainties), and are captured by a complex effective $\omega_r$ with $\text{Im}(\omega_r)$ of the order of $10^{-2}$. 
	This contrasts with uniformly capacitive circuits (all-cap, dark magenta), which exhibit logarithmic scaling with no non-monotonic peaks.
}
	\label{fig2}
\end{figure*}

\subsection{Violation of logarithmic impedance scaling}

To put our anomalous circuit impedance scaling behavior into perspective, we introduce the quantity $\beta = -\frac{d\log |Z|}{d\log N}$, which is the fractional rate of change of the impedance $Z$ with the system size $N$. It is closely related to the $\beta$-function in renormalization group analysis~\cite{callan_broken_1970, fisher_renormalization_1974, wilson_renormalization_1975}, and has also been famously employed in understanding the conductivity localization transition ~\cite{abrahams_scaling_1979, altshuler_interaction_1980, abrahams_quasiparticle_1981, lee_disordered_1985,garcia-garcia_anderson_2005,garcia-garcia_semi-poisson_2006} due to disorder scattering. 

In most conductors where $Z\sim N^{2-D}$, we have a constant $\beta = D-2$, which indicates that the impedance $Z$ increases (decreases) with the system size in a consistent qualitative manner for $D\leq2$ ($D>2$). 
This is the case for purely resistive media (such as the conducting plate depicted in Fig.~\ref{fig1}a) for which the impedance scales logarithmically viz. $Z\sim \log N$, giving rise to $\beta \sim -(\log N)^{-1}$ as sketched in Fig.~\ref{fig1}b (blue, green and dark green). However, we unveil that this crossover to the asymptotic limit can be far from smooth when a AC frequency scale exists in the circuit. As plotted in Fig.~\ref{fig1}b (red) for an illustrative 2D AC circuit lattice (detailed later), $\beta$ fluctuates erratically and dramatically as the system size $N$ increases. (Note that the irregular impedance scaling depicted in red in Fig.~\ref{fig1}b is not exclusive to a 2D sample but can occur in an $LC$ lattice, regardless of their dimensions.) 
In the following sections, this anomalous scaling behavior will be revealed to be part of an intricate fractal-like characteristic with slightly different reactance parameters often giving rise to unpredictably distinct anomalous impedance scaling.

\subsection{RLC Circuit with anomalous impedance scaling}

We investigate the discretization of the simplest 2D conducting sample, which is a $N\times N$ square lattice circuit array with fixed $RLC$ components connecting each node [Fig.~\ref{fig2}a]. For consistency, we shall always measure the impedance across two diagonally opposite corner nodes, even though the subsequent results remain qualitatively valid for arbitrary impedance intervals. If every connection in the square lattice is composed of the same element $z$, it can be shown (see Methods) that the corner-to-corner impedance scales like $Z\sim z\log N$ [Fig.~\ref{fig2}d]. This is not surprising, since it is only natural to expect that the square circuit lattice inherits the same logarithmic scaling as its continuum counterpart. 

Yet, we find that this usual logarithmic impedance scaling becomes severely violated when the lattice connections $z$ are replaced by two different circuit components with impedance of opposite signs, such as $L$ and $C$ components, which define a frequency scale. Specifically, we built an $N\times N$ square lattice circuit array on circuit boards [Fig.~\ref{fig2}b] ($N=2,..,7$), such that each horizontal link contains a capacitor $C$ and each vertical link contains an inductor $L$. In momentum space, the circuit Laplacian $\mathcal{L}$, which relates the voltage and input current profiles via $\bold I=\mathcal{L}\bold V$, takes the form
\begin{align}
\mathcal{L}(k_x,k_y)&=2i\omega C(1-\cos k_x) + \frac{2}{i\omega L}(1-\cos k_y)\notag\\
&=2i\omega C\left[(1-\cos k_x)-\omega_r^{-2}(1-\cos k_y)\right],
\label{circuit_Laplacian}
\end{align}
where $\omega$ denotes the AC driving frequency. Barring the $2i\omega C$ overall prefactor, $\omega_r=\omega\sqrt{LC}$ is the only nontrivial parameter of our circuit besides the lattice size $N$, neither of which constitutes another competing length scale. 

For a fixed $\omega_r$, the measured corner-to-corner impedance does not follow a simple trend with the lattice size $N$, but varies erratically with abrupt and prominent peaks at certain $N$. As plotted in Fig.~\ref{fig2}c for an ideal $LC$ circuit without any dissipation, the impedance $Z$ fluctuates wildly as $N$ is increased, such that $|Z|$ can abruptly become a few orders of magnitude larger for particular values of $N$. Besides, the impedance behaves qualitatively differently for different $\omega_r$, even across small changes in $\omega_r$. It is noteworthy that such ``quasi-random'' behavior is robustly measurable in an actual experimental implementation with inevitable dissipation, as reflected in our measured data (Fig.~\ref{fig2}d), which agrees well with theory despite unavoidable parasite and contact resistances as well as component disorder.

\subsection{Emergent fractal resonances}

\begin{figure}
	\includegraphics[width=\linewidth]{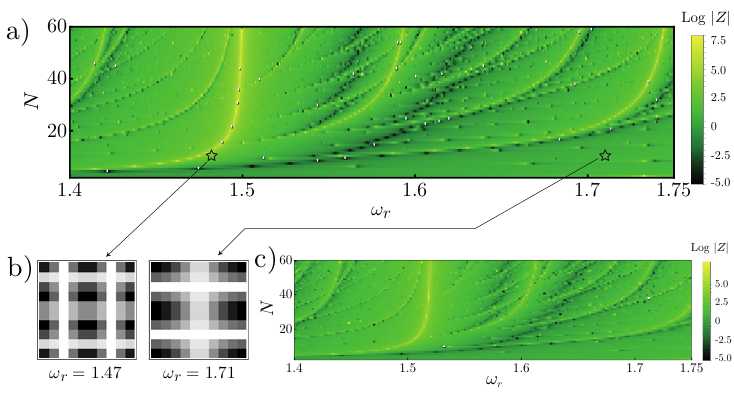}
	\caption{\textbf{Fractal nature of anomalous impedance scaling.} (a) Log-density plot of the corner-to-corner impedance $Z$ in the parameter space of (real) $\omega_r$ and lattice size $N$ showing variations of $Z$ across several orders of magnitude in the form of fractal-like branches. The ``branch'' near $\omega_r=1.5$ is the strongest but, still, it contains strong impedance peaks only for certain system sizes $N$. (b) Representative minimal-eigenvalue eigenstates of the Laplacian $\mathcal{L}$ with $N=9$ at two illustrative $\omega_r$ (Eq.~\eqref{resonance}) with very contrasting impedances $Z_{\omega_r=1.47}/Z_{\omega_r=1.71}=256/4.81\approx 53$. Unlike the case of waveguides, the markedly different impedances are not due to the spatial eigenstate distributions, which are qualitatively similar, but rather the ``vanishing energetics'' of $\omega_r$. (c) The impedance peak branches of $\log|Z|$ remain mostly robust in the presence of inevitable resistances, such as shown here for $\text{Im}(\omega_r)=-0.02$, which is of the same order as the parasitic resistances in our fabricated circuits. The plot legends in panels (a) and (c) indicate that the values represented are the logarithms of the absolute corner-to-corner impedance.}
	\label{fig3}
\end{figure}

The erratic, random-like behavior of the impedance across our $LC$ circuit suggests a hidden layer of emergent complexity in its resonance properties. Usually, one would expect a simple array of $LC$ components to behave as a waveguide with resonances that are simple enough to list down, for instance like the vibration modes on a stretched drumskin~\cite{kac1966can,tien1977integrated}. A complete impedance plot of our $LC$ circuit in $(N,\omega_r)$ parameter space, however, reveals a complicated fractal-like structure that bears resemblance to the energy bands in the Hofstadter butterfly~\cite{hofstadter_energy_1976, albrecht_evidence_2001, koshino2006hall, hunt2013massive}. In Fig.~\ref{fig3}a, we observe the following intricate hierarchy of impedance peaks: apart from some ``main'' branches, there exists a proliferation of less regular peaks that appear and disappear with the discreteness of $N$, akin to the cringes on the surface of a human palm. Additionally, these fractal patterns in our 2D $LC$ circuit are not confined to 2D instances, much like the Hofstadter butterfly, which is specific to 3D and quasi-1D systems~\cite{koshino_hofstadter_2001,koshino_phase_2002}.

To mathematically understand the origin of this fractal impedance behavior, we start from the formal expression for the impedance between two nodes $i$ and $j$~\cite{tzeng_theory_2006,lee_topolectrical_2018} 
\begin{align}
Z_{ij}&= \frac{V_i-V_j}{I}\notag\\
&=\frac{[\mathcal{L}^{-1}\bold I ]_i-[\mathcal{L}^{-1}\bold I]_j }{I}\notag\\
&=\sum_{\mu\neq 0} \frac{|\psi_\mu(i)-\psi_\mu(j)|^2}{\lambda_\mu},
\label{two-point-imp}
\end{align}
where $V_i$ and $V_j$ are respectively the voltage potentials at the current input sites $i$ and $j$ at which $I_l=I(\delta_{il}-\delta_{jl})$ is nonzero. Here $\psi_\mu$ and $\lambda_\mu$ are the corresponding eigenvectors and eigenvalues of the Laplacian, whose pseudoinverse is given by $\mathcal{L}^{-1}=\sum_{\mu\neq 0}\lambda_\mu^{-1}|\psi_\mu\rangle\langle\psi_\mu|$, where $\mu\neq 0$ indicates the omission of the uniform eigenvector corresponding to an overall voltage offset. 

Evidently, impedance peaks arise if there are eigenvalues $\lambda_\mu$ that are almost zero (not exactly zero, as they cannot perfectly vanish in a realistic circuit experiment). Such peaks have been featured as ``topolectrical'' resonances when the circuit band topology enforces topological zero modes~\cite{ningyuan2015time,imhof_topolectrical-circuit_2018, yang_observation_2020}. In our context, there is \emph{no} topological mechanism, and we proceed by deriving a compact albeit slightly complicated expression for the impedance $Z=Z_{ij}$ between the corner nodes $i$ and $j$, as detailed in ```Methods: Detailed derivation of the corner-to-corner impedance". The idea is to first consider the circuit under a doubled system with periodic boundaries where $\mu$ in Eq. \eqref{two-point-imp} now labels the momentum eigenmodes $k_{x}=\frac{2\pi m}{2N}$, $k_{y}=\frac{2\pi n}{2N}$, and next employ the method of images to enforce the vanishing of currents across the $N\times N$ open boundaries. In doing so, we obtain the impedance
\begin{equation}
Z(N)=\frac{2}{i\omega CN^2}\sum_{n+m\in\text{odd}}\frac{\cos \frac{n\pi}{2N}\cos \frac{m\pi}{2N}\cos \frac{(n+m)\pi}{2N}}{(1-\cos \frac{n\pi}{N})-\omega_r^{-2}(1-\cos \frac{m\pi}{N})}.
\label{Z}
\end{equation}
The denominator in Eq.~\eqref{Z} resolves the origin of incommensurability leading to fractal-like behavior. Analogous to the Harper equation describing a Landau level due to a magnetic field~\cite{harper_single_1955, hofstadter_energy_1976, krasovsky_bethe_1999,poshakinskiy_quantum_2021}, we find the relation
\begin{equation}
\omega_r^2=\omega^2 LC = \frac{1-\cos\frac{m\pi}{N}}{1-\cos\frac{n\pi}{N}}
\label{resonance}
\end{equation}
describing a circuit resonance. Here, $\omega_r^2$ plays the analogous role to the energy in the Hofstadter butterfly, and $N$ plays the role of the denominator defining a fractional flux. In our case, however, all rational fractions with denominator $N$ simultaneously contribute to the impedance, and a strong resonance occurs if there exist integers $m,n$ of the same parity that accurately satisfy Eq.~\eqref{resonance}.  

This relation explicitly expresses the resonance strength in terms of the commensurability properties of $\omega_r^2$ and $N$, even though the relation is hard to guess from intuitive reasoning. Unlike the Hofstadter butterfly problem, which is based on magnetic translation-symmetric Bloch states~\cite{hofstadter_energy_1976, koshino2006hall}, our circuit setup contains no such symmetries. While generic $LC$ (or likewise $RLC$) circuits do possess resonances, their resonance properties dimensionally depend on the frequency scale $LC$, and in general do not depend systematically on the system size. In our case, it is the mirror symmetry about the boundaries that fortuitously restores sufficient symmetry to give rise to an explicit, and hence also measurable, commensurability relation.

Stemming from the approximate solutions to Eq.~\eqref{resonance}, the impedance peaks are primarily manifestations of commensurate ``energetics'' that lead to a vanishing $\omega_r$, rather than special spatial mode configurations. To illustrate this point, illustrative near-resonant and off-resonant eigenmodes of $\mathcal{L}$ are plotted in Fig.~\ref{fig3}b. Note that the near-resonant eigenmodes do not exhibit any spatial distribution particularities that distinguish them from ordinary eigenmodes contributing to far lower impedances.

\subsection{Robustness of fractal impedance peaks and crossover from logarithmic scaling}


In actual experiments, contact and parasitic resistances introduce inevitable dissipation and attenuate the impedance peaks, as evident in the comparison between Figs.~\ref{fig2}c and \ref{fig2}d. Yet, the key anomalous fractal scaling behavior of the impedance remains robust. Phenomenologically, we can represent these dissipative effects through modifying the capacitor and inductor impedances to $(i\omega C)^{-1} \rightarrow (i\omega C)^{-1} +R_C$ and $i\omega L \rightarrow i\omega L +R_L$, where $R_C$ and $R_L$ are real effective resistances. Incorporating the estimated $R_C$ and $R_L$ values from our experiments, which add an imaginary part of order $\mathcal{O}(10^{-2})$ to $\omega_r$, we find that the fractal parameter space profile of the impedance $|Z|$ becomes slightly smoothed out (Fig.~\ref{fig3}c), even though the main branches of the fractal structure remains qualitatively unchanged. This robustness stems from the strong impedance divergence due to commensurability effects on a discretized conducting medium, which holds for generic lattice discretizations, and not just for our square lattice (Eq.~\eqref{Z}).

\subsection{Anomalous impedance in the 2D honeycomb lattice}
\begin{figure}
	\includegraphics[width=1.0\linewidth]{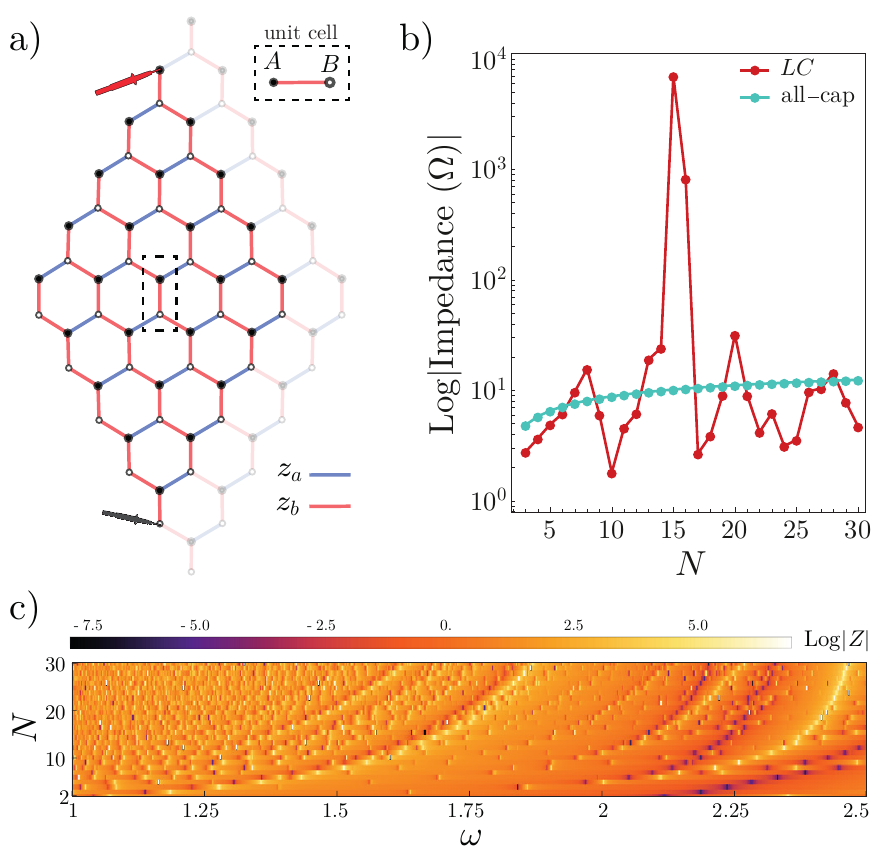}
	\caption{\textbf{Honeycomb circuit lattice and its impedance results.} (a) Illustrative honeycomb lattice with zigzag edge design when $N=5$. A unit cell consists of nodes belonging to two sublattices $A$ (black circles) and $B$ (black framed white circles). The blue and red lines represent the node links with the different admittances of $z_a$ and $z_b$, respectively. The faded cells indicate the extension of the circuit when $N=6$, as an example. (b) The impedance response of the circuit in (a). The circuit is a resonant medium when $z_a=1(i\omega L)$ and $z_b=i\omega C$ and presents sharp impedance resonances as a function of the circuit size. The parameters used are $\omega C=1$ for the uniform circuit made of only capacitors and $\omega C=1$, $\omega L=2.21$ for the $LC$ honeycomb circuit. (c) Fractal scaling of the 2D honeycomb lattice in the circuit size and driving frequency domain when $C=L=1$. The brightest and darkest branches represent the strong anomalous impedance resonances, which depend on the circuit size $N$. The legend located above the density plot indicates the logarithm of the absolute impedance.}
	\label{fig4}
\end{figure}
The presence of anomalous impedance in $LC$ circuits is a universal property that arises when two distinct components with opposing phases are present in the circuit lattice. Here, we investigate the impedance characteristics of the two most distant nodes in a honeycomb lattice with a zigzag edge design. We consider nodes belonging to two sub-lattices $A$ and $B$ that are connected by an admittance $z_a$ (blue lines in Fig.~\ref{fig4}a). The nearest neighbor nodes are also connected by an admittance $z_b$ (red lines in Fig.~\ref{fig4}a) in a unit cell. The resultant lattice is fully reactive and non-resonant when $z_a$ and $z_b$ have the same phases and resonant when they have opposite phase. To investigate the lattice size-dependent impedance characteristics, we examine the circuit in both cases and find that there are anomalous impedance resonances at specific circuit sizes when the circuit parameters are fixed. Fig.~\ref{fig4}b illustrates the impedance across circuit size behavior under two conditions: when the entire circuit is composed of only a single type of capacitor with a capacitance of $z_a=z_b=i\omega C$, and when it is composed of two distinct admittances of $z_a=1/(i\omega L)$ and $z_b=i\omega C$. The fractal scaling observed in the 2D $LC$ circuit (Fig.~\ref{fig2}a) also arises in the 2D honeycomb lattice. Fig.~\ref{fig4}c illustrates the impedance resonances exhibiting fractal-like scaling with the variation of the circuit size and driving frequency. Furthermore, the edge design of the lattice affects only the form of the fractal-like branches, but the branches persist across different edge designs. This validates the fractal-like anomalous impedance scaling in $LC$ circuits with different lattice models.

\section{Discussion}

In this work, we theoretically and experimentally investigated the pronounced yet seemingly random impedance scaling behavior of $RLC$ circuit lattices arrays. This anomalous impedance scaling contrasts greatly with the usual logarithmic scaling expected in 2 dimensions, and is rooted in the commensurability properties of the circuit Green's function eigenvalues, reminiscent of the commensurability conditions pertaining to a Hofstadter lattice with magnetic flux. This results in a curious fractal-like impedance behavior in the parameter space of dimensionless frequency $\omega_r$ and lattice size $N$, whose complexity and structure elude any simplistic waveguide analysis.

In generic circuit lattices with more complex connections, unit cells, and feedback elements, more sophisticated fractal impedance fringes would be expected due to the more complicated commensurability conditions for the vanishing of the circuit Laplacian eigenvalues. This points towards the hitherto unnoticed general breakdown of a continuum description of resonant conducting media, which highlights the need for more careful analysis in the discretization of device geometries in electrostatics simulations. The discretization of continuous media involves dividing the medium into discrete units or elements that can be represented using discrete variables. In the context of electrical circuits, this can involve dividing continuous electrical fields or currents into discrete components such as resistors, capacitors, and inductors, which can be connected in various ways to create a circuit. Discretization allows for the use of mathematical tools and techniques to analyze physical phenomenon in an continuum media~\cite{webman_theory_1977,kirkpatrick_percolation_1973}, as in this study.

More generally, the fractal anomalous scaling behavior extends to the steady state behavior of systems governed by network Laplacians where a resonance frequency also enters the dynamics. This includes, for instance, directed information networks, which are physically unrelated to electrical circuits. While we have focused on a very regular square lattice network that should have possessed simple logarithmic impedance scaling naively, such fractal scaling also exists in more generic network structures, albeit in possibly more concealed manners.

\section{Methods} 

\subsection{Detailed derivation of the corner-to-corner impedance}

Here, we provide a detailed pedagogical exposition of the impedance formula in Eq.~\eqref{Z}, which has an analytic expression thanks to the fact that the method of images can be used in implementing the circuit boundaries. Besides our method, one can also compute the complex equivalent resistance in finite complex lattices by using recursion-transform method~\cite{tan_recursion-transform_2015,tan_recursion-transform_2015-2,tan_recursion-transform_2017}.

The impedance is a measure of the total resistance that a circuit offers to the flow of alternating current. It is composed of reactance, which is the resistance of a circuit element to AC due to its inherent capacitance or inductance, and resistance, which is the resistance of a circuit element to AC due to its inherent properties (i.e., impedance $Z = X_C + R + X_L$, where $X_C= 1/(i\omega C)$ and $X_L = i\omega L$ are the reactances corresponding to a capacitance $C$ and inductance $L$, respectively, and $R$ is the resistance).

\begin{figure}[h]
	\centering
	\includegraphics[width=1\linewidth]{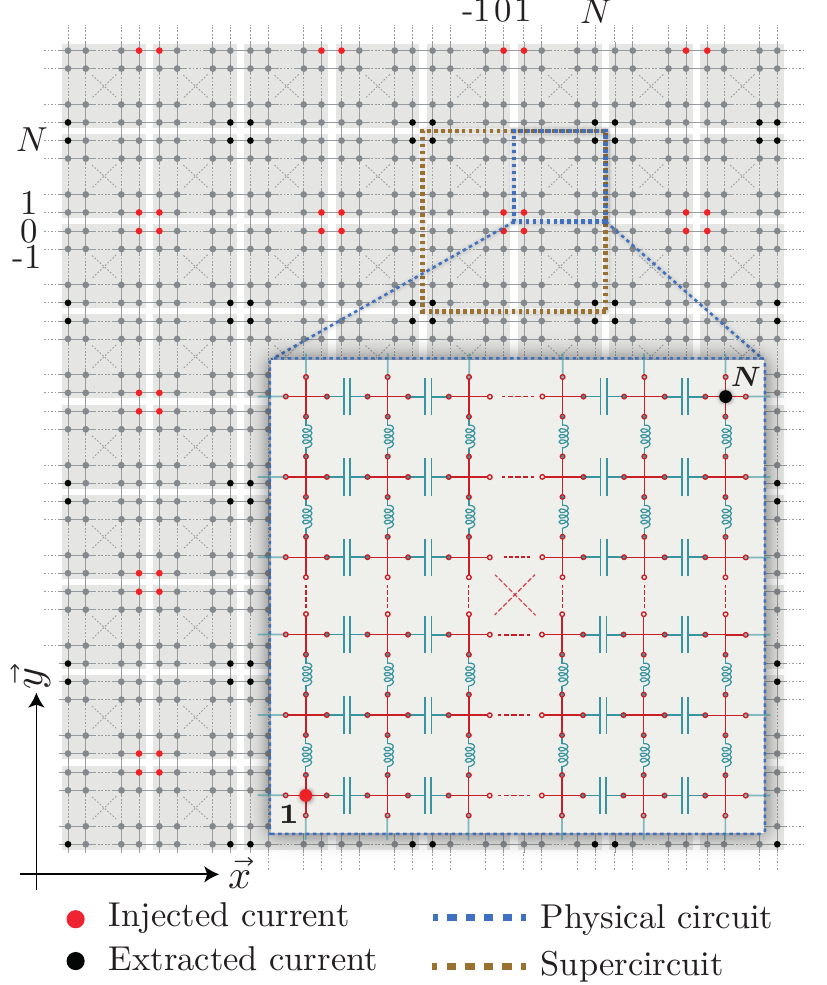}
	\caption{\textbf{Uniform infinite lattice tiling with image copies of our 2D $LC$ circuit.} Red and black dots signify the nodes where current is injected and extracted, respectively. The blue dashed square represents the physical $N$-unit cell 2D $LC$ circuit, while the brown dashed square illustrates the $2N$-unit cell supercircuit, which includes the image copies of the $N$-unit cell circuit along each principal direction. }
	\label{fig5}
\end{figure}

Any two-point impedance can be derived from Eq.~\eqref{two-point-imp}, which requires determining the electric potential difference between two nodes in response to a current $I$ injected at node $i$ and extracted at node $j$ \cite{tzeng_theory_2006}. To derive the analytical corner-to-corner impedance expression for a finite $N \times N$ square circuit, we consider an infinite circuit network built up from copies of the original $ N \times N$ circuit (see Fig.~\ref{fig5}). (Here, $N$ is the circuit size in terms of the number of unit cells.) We then label the nodes in this infinite circuit lattice such that the diagonally opposite corner nodes of the physical single finite $N\times N$ circuit are  $\mathbf{1}$ and $\mathbf{N}$, where $\mathbf{1}=\mathbf{a}_1+\mathbf{a}_2$, $\mathbf{N}=N\mathbf{a}_1 + N \mathbf{a}_2$ and $\mathbf{a}_1$ and $\mathbf{a}_2$ are the unit vectors for a 2-dimensional circuit. 
In such an infinite network, the current radiating from node \textbf{r} can be viewed in analogy to the flow of a current density $\mathbf{j}$ through a material with conductivity $\sigma$ in response to the electric field $\mathbf{E}$ resulting in current distribution $\mathcal{I} = \div{(\sigma \mathbf{E})} $. Since $\mathbf{E} = -\grad V$ where $V$ is the electric potential, we arrive at the Poisson's equation
\begin{equation}
    \nabla^2 V = - z \mathcal{I},
    \label{Poisson_eq}
\end{equation}
where $\sigma^{-1}$ in the continuum system is now replaced by $z$, the coupling impedance between each neighbor in the circuit. To solve Eq.~\eqref{Poisson_eq} for a single copy of the $N\times N$ circuit array, we invoke the method of images, through which the electric potential $V$ in a specific region with boundaries can be obtained. In the context of electrostatics, a classic application of the method of images is to solve for the electric potential on a conducting plate stemming from a point charge. To achieve this, one can replace the conducting plate by an image charge of opposite sign located spatially opposite to the original charge \cite{riley_mathematical_1999,griffiths_introduction_2005,jackson_classical_1999,yang_designing_2022}. By using the same perspective, we can regard the finite $N \times N$ circuit as the analog of the original source and its boundaries as conducting equipotential plates across which no current can flow~\cite{mamode_electrical_2017}. Applying the method of images, the $N \times N$ circuit is replicated to form a 2D infinite network with capacitive $C$ and inductive $L$ couplings along the horizontal and vertical links, respectively (refer Fig. \ref{fig5}). The Poisson equation in Eq.~\eqref{Poisson_eq} is solved for this infinite 2D network. The solution for the node potentials within the boundaries of the source $N \times N$ plate will correspond to the node potentials of the original finite circuit. This is because the presence of the replicas, which serve as the analogues of the image charges, ensures that the potential differences between all the edge nodes along the entire boundaries of the `source' network and those of its image copies are zero. Hence, no current would flow across the edges of the $N \times N$ source network, which is the exact boundary condition for the original finite $N \times N$ circuit. To implement the method of images in the circuit network, we write the current distribution over the supercircuit with the period of $2N$ as 
\begin{equation}
	\mathcal{I}(\mathbf{r})=I\big(\delta(\mathbf{r},\mathbf{r}_\text{in})-\delta(\mathbf{r},\mathbf{r}_\text{out})\big),
\end{equation}
where $\mathbf{r}_{\text{in}}=\{(0\mathbf{a_1},0\mathbf{a_2}),(1\mathbf{a_1},0\mathbf{a_2}),(0\mathbf{a_1},1\mathbf{a_2}),(1\mathbf{a_1},1\mathbf{a_2})\}$, $\mathbf{r}_{\text{out}}=\{(N\mathbf{a_1},N\mathbf{a_2}),(-(N-1)\mathbf{a_1},N\mathbf{a_2}),(N\mathbf{a_1},-(N-1)\mathbf{a_2}),(-(N-1)\mathbf{a_1},-(N-1)\mathbf{a_2})\}$. Here the vector $\mathbf{r}$ for our 2D circuit is defined as $\mathbf{r}= n \mathbf{a}_1+ m \mathbf{a}_2$ where $(n,m)$ are integers varying between $-(N-1)$ to $N$ and $\delta$ represents the Kronecker delta rather than the Dirac delta function of the continuum electrostatic model, due to the discrete locations of the injected/extracted currents (at the circuit nodes) and their finite magnitudes. Due to the inversion and translation symmetries of our circuit, the potential distribution produced by the current distribution can be obtained by translating the extracted currents by any multiple of the $2N$ period, i.e., $\mathbf{r}_{\text{out}}=\{(N\mathbf{a_1},N\mathbf{a_2}),((N+1)\mathbf{a_1},N\mathbf{a_2}),(N\mathbf{a_1},(N+1)\mathbf{a_2}),((N+1)\mathbf{a_1},(N+1)\mathbf{a_2})\}$ (see Fig. \ref{fig5}). Therefore, the voltage potential in Eq.~\eqref{Poisson_eq} by means of the definition of the Green's function $\delta (\mathbf{r}) = - \nabla^2 G(\mathbf{r})$ and by considering all the image current injection/extraction points can be found as
\begin{equation}
	\begin{aligned}
    V (\mathbf{r}) = z I\big(&G(\mathbf{r})+G(\mathbf{r}+\mathbf{a}_1)+G(\mathbf{r}+\mathbf{a}_2)+\\&
    G(\mathbf{r}+\mathbf{a}_1 + \mathbf{a}_2) -G(\mathbf{r+N})-\\
    &G(\mathbf{r}+\mathbf{N} + \mathbf{a}_1)-G(\mathbf{r}+\mathbf{N} + \mathbf{a}_2)-\\
 	&G(\mathbf{r + N}+\mathbf{a}_1 + \mathbf{a}_2) \big).
   	\end{aligned}
    \label{potential_at_r}
\end{equation}
\noindent  Now that we have reformulated the finite circuit problem as a problem on an infinite 2D lattice, we can find the momentum space Green's function by employing the discrete Fourier transform (recalling $G(\mathbf{k})=\mathcal{L}^{-1}(\mathbf{k})$):

\begin{equation}
    G(\mathbf{r}) = \frac{1}{(2 N)^2} \sum_\mathbf{k} \frac{e^{i\mathbf{r}.\mathbf{k}}}{\mathcal{L}(\mathbf{k})},
    \label{Greens_func_k}
\end{equation}

\noindent where $\mathcal{L}(\mathbf{k})$ is the corresponding circuit Laplacian given in Eq.~\eqref{circuit_Laplacian}. Here, the momentum space vectors for our 2D circuit are $\mathbf{k}=k_{x}\mathbf{a_1}+k_{y}\mathbf{a_2}$ with $k_{x}= 2n \pi /2N$ and $k_{y}= 2m \pi /2N$, where $n$ and $m$ are integer momentum indices from $1$ to $2N$. Because current is injected and extracted at opposite diagonal corners of the circuit and considering the translational invariance of the infinite circuit lattice, by symmetry \cite{cserti_application_2000,atkinson_infinite_1999,cserti_uniform_2011}, $V(\mathbf{1}) = - V(\mathbf{N})$. Thus, $Z/I=V(\mathbf{1})-V(\mathbf{N})=2V(\mathbf{1})=-2V(\mathbf{N})$. From the momentum space Green's function of Eq.~\eqref{Greens_func_k} and making use of Eq.~\eqref{potential_at_r},
\begin{equation}
\begin{aligned}
  &V(\mathbf{1})=-V(\mathbf{N})= \frac{I}{4 N^2} \sum_{n} \sum_{m} \times\\
     & \frac{(1-(-1)^{n+m})(1+e^{i \pi n/N}+e^{i \pi m/N}+e^{i \pi (n+m)/N})}{i\omega C (1-\cos(n\pi /N))+\frac{1}{i\omega L}(1-\cos (m\pi /N))}.\\
\end{aligned}
\label{Elec_pot_final_exp}
\end{equation}

\noindent Here, the summation is taken over $(n+m)\in$ odd because $(1-(-1)^{n+m})$ in the numerator is zero when $(n+m)$ is even and $2$ when $(n+m)$ is odd. Notice that the unit impedance $z$ between the couplings in Eq.~\eqref{potential_at_r} is now replaced by the impedances $i\omega C$ and $1/(i\omega L)$ in the Laplacian in the denominator of Eq.~\eqref{Elec_pot_final_exp}. We convert the numerator of Eq.~\eqref{Elec_pot_final_exp} to the trigonometric form by using the identity $1+\cos(A)+\cos(B)+\cos(A+B)=4\cos(A/2)\cos(B/2)\cos((A+B)/2)$. After the transformation, the imaginary part of the expansion (i.e., the $i\sin(\mathbf{k}_{i})$ terms) disappears because it cancels out. 
The impedance between two opposite corners (i.e., node $\mathbf{1}$ and node $\mathbf{N}$) in our 2D $N\times N$ $LC$ circuit is then obtained as

\begin{equation}
Z(N)=\frac{2}{i\omega C N^2} \sum_{n=1}^{2N}\sideset{}{^*} \sum_{m=1}^{2N}\frac{\cos \frac{n\pi}{2N}\cos \frac{m\pi}{2N}\cos \frac{(n+m)\pi}{2N}}{(1-\cos \frac{n\pi}{N})-\omega_r^{-2}(1-\cos \frac{m\pi}{N})},
\label{two-point-Z}
\end{equation}

\noindent where $Z(N)$ represents the two-point impedance between the two diagonally opposite nodes as a function of the circuit size $N$ and $\omega_r=\omega \sqrt{LC}$. The asterisk indicates that the summation should be taken only over odd values of $(n+m)$ where $(n,m) \in \{1,2,...,2N\}$. Note that the summation can be performed over $2N$ period due to the translation invariance. The condition for a vanishing denominator, i.e. divergent $Z(N)$, is exactly that of the resonance condition given by Eq.~\eqref{resonance}.

\subsection{Determination of the fractal dimension}
\begin{figure}[h]
	\centering
	\includegraphics[width=7cm]{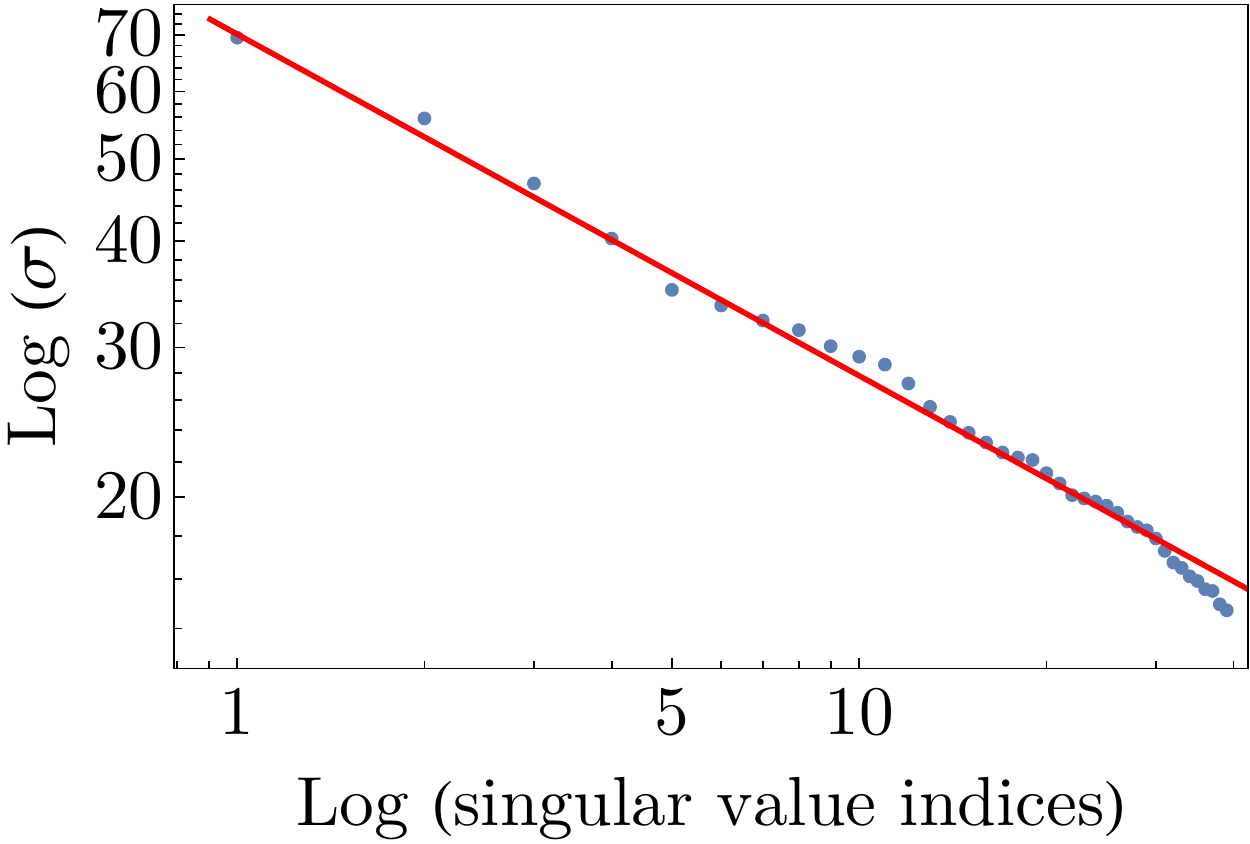}
	\caption{\textbf{Log-log scaling plot of the singular values extracted from the fractal diagram of our 2D circuit, obtained via Singular Value Decomposition (SVD).} The SVD is performed on the fractal matrix $\log |Z|$ displayed in Fig.~\ref{fig3}a. The blue dots represent the singular values given by Eq.~\eqref{svd}, while the red solid line represents the best linear fit of the data. The fractal dimension is calculated as one minus the slope of the linear fit, resulting in $D=1.4$ where $D$ represents the fractal dimension.}
	\label{fig6}
\end{figure}
It is possible to estimate the fractal dimension $D$ by using the Singular Value Decomposition (SVD) method~\cite{weng_singular_2022,alter_singular_2000,lee_exact_2015}. In this method, the self-similarities~\cite{mishra_effective_2021} or fractal dimension in a dataset, which is the fractal diagram in Fig.~\ref{fig3}a  (i.e., $\log|Z|$), is given by one minus the slope of the log-log plot of the singular values of the fractal matrix~\cite{malcai_scaling_1997,carr_practice_1991}. To determine the fractal dimension of our fractal structure, we performed SVD and write the decomposed matrices as
\begin{equation}
	\log|Z|=u~\sigma~v^\intercal
	\label{svd}
\end{equation}
where ($^\intercal$) denotes the transpose operation, $u$ and $v$ are the left and right singular matrices, respectively, and $\sigma$ is a diagonal matrix that comprises the singular values of the fractal. These diagonal values are nonnegative, and their squares give the eigenvalues of the $\log|Z|$ matrix~\cite{alter_singular_2000,weng_singular_2022}. We arrange the singular values in decreasing order, such that the largest value is $\sigma_1$, the second-largest is $\sigma_2$, and so on, i.e., $\sigma_1>\sigma_2>\sigma_3>\cdots$. This allows us to determine the scaling ratio, which is defined as the ratio of the largest singular value to the fractal matrix dimension. For example, in our case in Fig.~\ref{fig3}a, we find that $\sigma_1=276.71$ and $\text{dim}(\log|Z|)= 388$, and thus determine the scaling ratio as $\sim0.7$. The fractal dimension can be defined by utilizing the slope $(m_s=-0.4)$ of the best linear fit in the log-log scale plot (Fig.~\ref{fig6}) ~\cite{faloutsos_fast_2021}. Therefore, the fractal dimension is determined as $D=1-m_s=1.4$.

\subsection{Effects of inevitable parasitic resistances}
\begin{figure}[h]
	\centering
	\includegraphics[width=\linewidth]{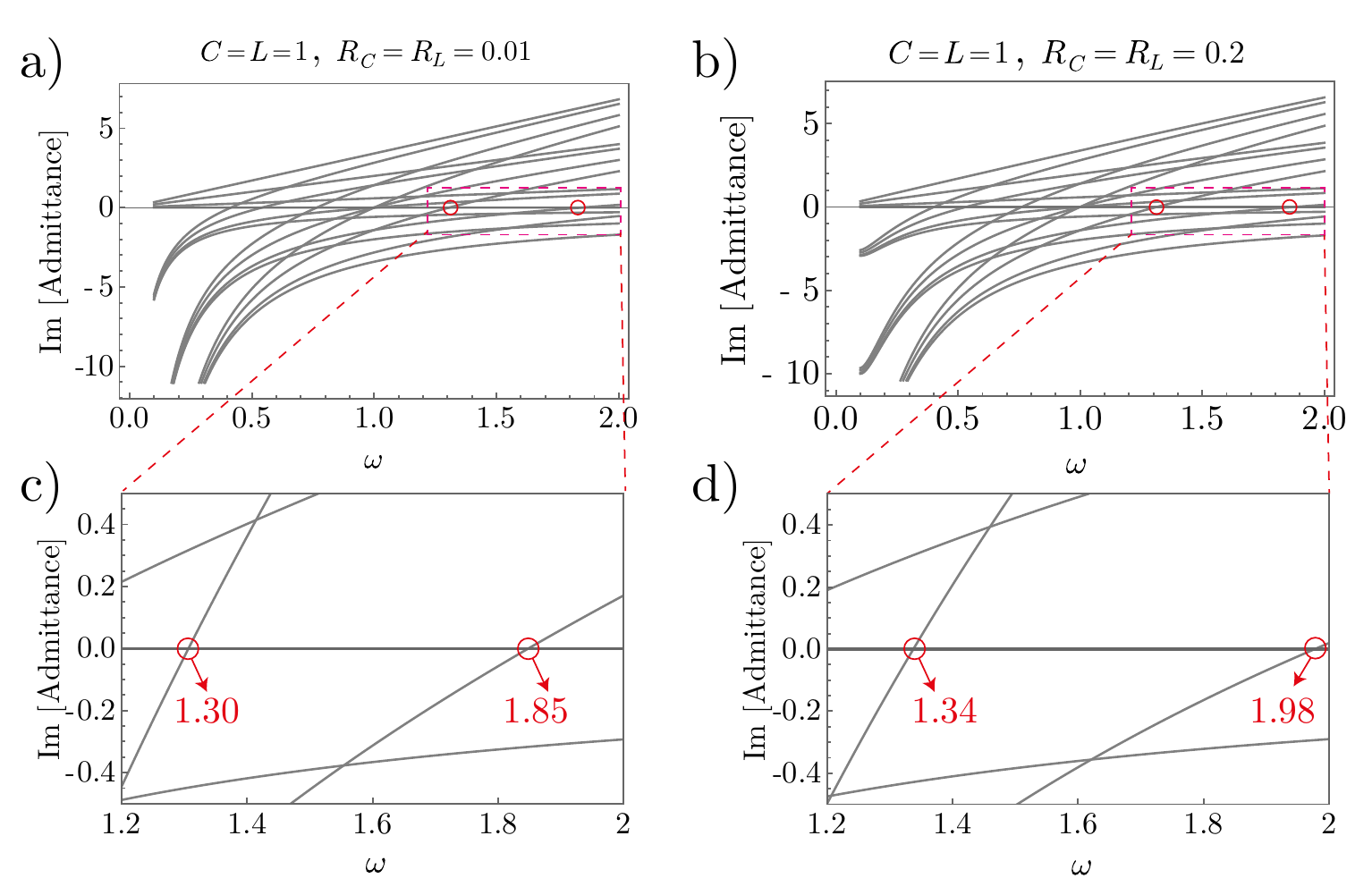}
	\caption{\textbf{Panels (a) and (b) show the admittance spectra of the 2D $LC$ square circuit when $C=L=1$ and $N=3$.} The upper row shows the admittance spectrum under the consideration of the parasitic series resistances when $R_C=R_L=0.01 \Omega$ and $R_C=R_L=0.2 \Omega$, respectively. Panels (c) and (d) display zoomed-in views of points where two randomly chosen admittance bands, depicted in (a) and (b) respectively, intersect the zero-admittance axis. An increase in parasitic resistances results in a shift towards higher frequencies at the band crossing points.}
	\label{fig7}
\end{figure}
Theoretically, the parasitic resistances can be incorporated by introducing additional the real effective resistances $R_C$ for the capacitors and $R_L$ for the inductors, as we discussed in the main text. In Fig.~\ref{fig3}c, we plot the fractal impedance peaks under the consideration of these parasitic resistances, which contribute to the imaginary part of the circuit resonance condition ($\omega_r$). As can be seen in comparison with Fig.~\ref{fig3}a, the realistic parasitic resistances only lead to a smooth shift of the impedance peak branches while the size-dependent resonances persist without further losses. This behavior can be understood from the admittance band structure. Since all the information for an ideal fully resonant media is contained within the imaginary part of its impedance, the presence of the parasitic resistances makes the impedance complex. This complexity implies that the stored energy represented by the imaginary part is dissipated due to the parasitic resistances. However, as long as the parasitic resistances do not become dominant, their presence results in a smooth shift in the frequency corresponding to zero admittance in the admittance band structure. To show this, we plot the admittance band structures in Fig.~\ref{fig7}a, and \ref{fig7}b for our 2D $LC$ circuit with $N=3$ by considering two different parasitic resistances. As evident from Figs.~\ref{fig7}c and ~\ref{fig7}d, which provide a zoomed-in view of the band crossing points in panels (a) and (b) respectively, the admittance bands themselves remain qualitatively unchanged, although there is a shift towards higher frequencies in the admittance band structure. This is significant because impedance resonances occur in the presence of nearly zero eigenvalues, which correspond to the band-crossing points in the Laplacian formalism. According to Eq.~\eqref{two-point-imp}, a large impedance is obtained when at least one of eigenvalues ($\lambda_\mu$) becomes nearly zero provided that the wavefunction values at the measurement points of its corresponding eigenstate are not zero. Fig.~\ref{fig7} demonstrates that introducing small parasitic resistances results in a shift in the frequencies corresponding to the zero-energy eigenvalues. This explains the shift to the right in Fig.~\ref{fig3}c and demonstrates the robustness of our circuit against parasitic resistances despite the smooth shift in the resonant frequencies.

\subsection{Details of the experiment}

Our experiment consists of measuring the corner-to-corner impedance of a square lattice array of $LC$ elements, as pictured in Fig.~\ref{fig2}a, b. To fit our measured impedances with the theoretical predictions, we introduce serial resistances to the $L$ and $C$ components, such that the effective $\omega_r$ becomes complex. Below, we detail the procedures involved, as well as some of the subtleties.

\begin{figure}
	\centering
	\includegraphics[width=\linewidth]{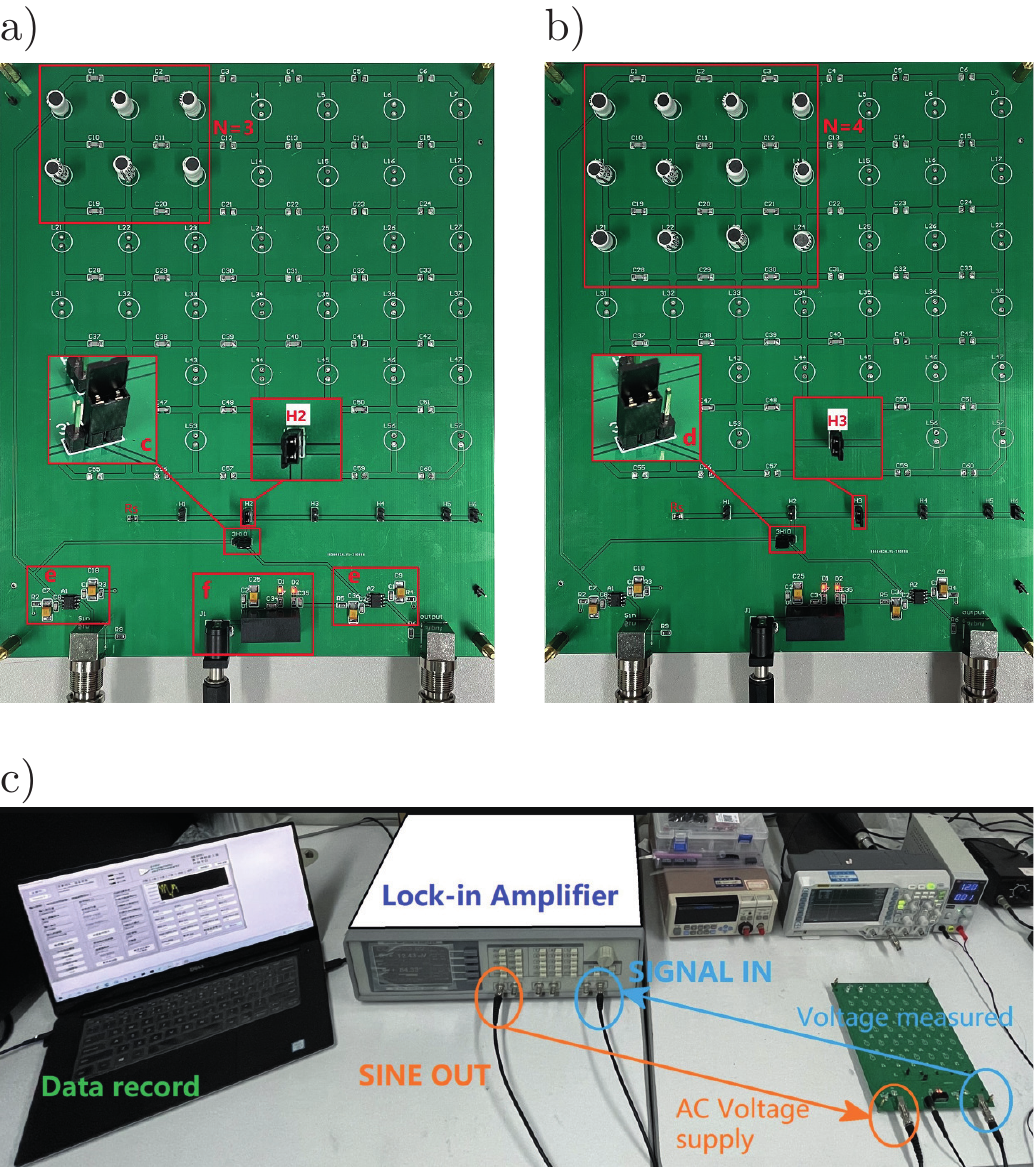}
	\caption{\textbf{Methodology of measuring and extending the $N\times N$ circuits. } (a) For the $N=3$ case measurement, H2 was connected through the jumper cap to connect the entire measurement circuit after the components were soldered. The voltage across a standard resistor of 110 $\Omega$ was then measured by connecting the right side of the switch 3H10 (c in Fig.(a)). The voltage across the entire circuit was next measured using lock-in amplifier by connecting the left side of the switch 3H10 with a jumper cap (d in Fig.(b)). After the measurement for $N=3$ was completed, all the switches were disconnected, and the $N=4$ circuit  (b) was extended from the $N=3$ circuit. H3 was subsequently connected and the above steps were repeated after the additional required components were soldered. In addition to the measured circuit, an operational amplifier (e in Fig.(a)) was also added at the input end and another operational amplifier at the output end of the signal as followers to ensure the stability of the lock-in amplifier signal. The power supply module (f in Fig.(a) ) supplies power to these two operational amplifiers. 
		(c) Lab setup. To effectively avoid interference to the weak signals, we used a lock-in amplifier for measurements. SINE OUT provides an AC voltage signal to the measurement circuit, and SIGNAL IN measures the voltage across either a standard resistor or the entire circuit (controlled by a switch). 
	}
	\label{figExp}
\end{figure}

\subsubsection{Measurement process}

Our measurements were performed on circuit lattices of different sizes corresponding to $N=2$ to $7$ (refer to Fig.~\ref{figExp}a and ~\ref{figExp}b). To mitigate the effects of component disorder, the larger lattices were built by extending the smaller lattices, i.e., measurements are carried out on a circuit of size $N$ before the circuit was extended by soldering additional circuit elements to form a circuit of size $N+1$. 

Based on the fractal parameter space diagram, two  AC frequency ranges of interest were determined as 115-175 kHz and 215-290 kHz. For each lattice size $N$, we swept through both of these ranges with a sweep step size 200 Hz (an overly small step size will significantly increase the measurement time). The sweep time interval was set to 1000 ms, which was sufficient to ensure that the voltage reached a stable state each time the frequency $\omega$ was updated. For each $(\omega,N)$ point, the last three (stabilized) voltages were averaged and recorded. The configuration of the measurement setup is shown in Fig.~\ref{figExp}c.


\begin{figure}
	\centering
	\includegraphics[width=\linewidth]{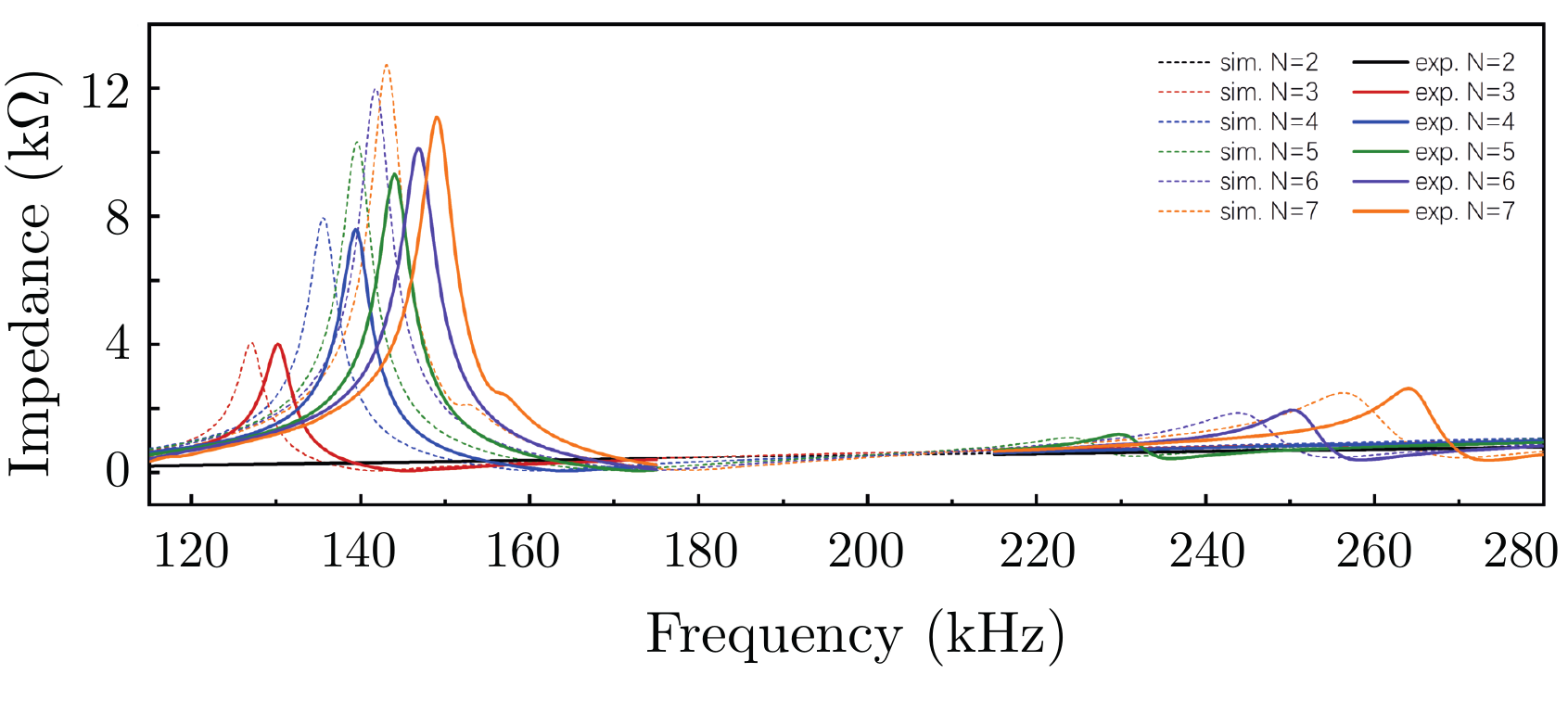}\\
	\caption{\textbf{Comparison of theoretically simulated and experimentally measured impedances}. The simulated and experimentally measured impedances differ slightly in both the peak positions and heights, but the discrepancies are fully accounted for by component uncertainly and tolerances as detailed in Table~\ref{table1}. }
	\label{fig9}
\end{figure}

\subsubsection{Analysis of uncertainties}
There are two main types of discrepancies between the theoretically predicted (Eq.~\eqref{Z}) and experimentally measured impedances. The first is the discrepancy between the predicted and measured resonant frequencies \emph{f$_0$} where the impedance peaks, and the second is the discrepancy between the predicted and measured impedance values at \emph{f$_0$}. The first discrepancy can mainly be attributed to the uncertainties in the component values. The components we used are rated at $ C=4.7\ \mathrm{nF}\pm1\% $, $ L=1\ \mathrm{mH}\pm5\% $. Employing the frequency scale $\omega/\omega_r=(LC)^{-1/2}$ as a value estimator, we find that the discrepancy of \emph{f$_0$} presented in Fig.~\ref{fig9} is within a reasonable range, as further tabulated in Table~\ref{table1}.
\begin{table*}[]
	\begin{tabular}{|c|ll|ll|ll|ll|}
		\hline
		\multirow{3}{*}{\textbf{\begin{tabular}[c]{@{}c@{}}\end{tabular}}} & \multicolumn{2}{c|}{\multirow{2}{*}{\textbf{\begin{tabular}[c]{@{}c@{}}Sim. with   \\ C=4.7nF$\pm$ 1\%, L=1mH$\pm$ 5\%\end{tabular}}}} & \multicolumn{2}{c|}{\multirow{2}{*}{\textbf{\begin{tabular}[c]{@{}c@{}}Sim. with   \\ C=4.7nF, L=1mH\end{tabular}}}} & \multicolumn{2}{c|}{\multirow{2}{*}{\textbf{Exp.}}}                          & \multicolumn{2}{c|}{\multirow{2}{*}{\textbf{Error}}}               \\
		& \multicolumn{2}{c|}{}                                                                                                       & \multicolumn{2}{c|}{}                                                                                               & \multicolumn{2}{c|}{}                                                        & \multicolumn{2}{c|}{}                                              \\ \cline{2-9} 
		& \multicolumn{1}{l|}{\textbf{range of $f_0(kHz)$}}                           & \textbf{range of $Z(k\Omega)$}                          & \multicolumn{1}{c|}{\textbf{$f_0(kHz)$}}                     & \multicolumn{1}{c|}{\textbf{$Z(k\Omega)$}}                    & \multicolumn{1}{c|}{\textbf{$f_0(kHz)$}} & \multicolumn{1}{c|}{\textbf{$Z(k\Omega)$}} & \multicolumn{1}{c|}{\textbf{$f_0$}} & \multicolumn{1}{c|}{\textbf{$Z$}} \\ \hline
		\textbf{3}                                                                       & \multicolumn{1}{l|}{123.4$\sim$131.0}                                    & 3.9$\sim$4.2                                     & \multicolumn{1}{l|}{127.0}                                & 4.06                                                    & \multicolumn{1}{l|}{130.2}            & 4.01                                 & \multicolumn{1}{l|}{2.52\%}      & -1.23\%                         \\ \hline
		\textbf{4}                                                                       & \multicolumn{1}{l|}{131.7$\sim$139.8}                                    & 7.6$\sim$8.2                                     & \multicolumn{1}{l|}{135.6}                                & 7.96                                                    & \multicolumn{1}{l|}{139.4}            & 7.59                                 & \multicolumn{1}{l|}{2.80\%}      & -4.65\%                         \\ \hline
		\textbf{5}                                                                       & \multicolumn{1}{l|}{135.5$\sim$144.0}                                    & 9.9$\sim$10.7                                    & \multicolumn{1}{l|}{139.6}                                & 10.32                                                   & \multicolumn{1}{l|}{144.0}            & 9.31                                 & \multicolumn{1}{l|}{3.15\%}      & -9.79\%                         \\ \hline
		\textbf{6}                                                                       & \multicolumn{1}{l|}{236.8$\sim$251.4}                                    & 1.8$\sim$1.9                                     & \multicolumn{1}{l|}{243.8}                                & 1.86                                                    & \multicolumn{1}{l|}{250.4}            & 1.95                                 & \multicolumn{1}{l|}{2.71\%}      & 4.84\%                          \\ \hline
		\textbf{7}                                                                       & \multicolumn{1}{l|}{248.9$\sim$264.2}                                    & 2.4$\sim$2.6                                     & \multicolumn{1}{l|}{256.3}                                & 2.48                                                    & \multicolumn{1}{l|}{263.8}            & 2.62                                 & \multicolumn{1}{l|}{2.93\%}      & 5.65\%                          \\ \hline
	\end{tabular}
	\caption{Comparison of theoretical simulation results with given component error tolerances against experimental results. The given $LC$ components are rated at $L=1\ \mathrm{mH}\pm 5\%$ and $C=4.7\ \mathrm{nF}\pm 1\%$. $Z$ and $f_0$ are the peak impedance and the frequency at which it occurs. The experimentally measured (exp.) values are indeed within the theoretically predicted ranges (sim.) corresponding to the error tolerances. The full exp. and sim. impedance curves given in Fig.~\ref{figzf}. 
	}
	\label{table1}
\end{table*}

The second type of discrepancy, i.e., the impedance peak heights, is greatly affected by the parasitic resistance in addition to the component uncertainties. The parasitic resistance effectively suppress the peak of the measured impedance. This is reflected in the impedance-frequency curve in which the decrease in the peak value is accompanied by an increase in the FWHM (full width at half maximum), which makes it difficult to distinguish between the curves of different system sizes if the parasitic resistances were too large (fortunately, they were not). There may be several sources that contribute to parasitic effects, such as parasitic resistance, capacitance, and inductance. However, through numerous simulation studies, we found that the parasitic resistances are the most significant contributors that affect the measured impedance resonances. Using the estimated serial parasitic resistances of $R_{pL} = 3.3 \Omega$, $R_{pC} = 4.5 \Omega$, $R_{pW} = 0.1 \Omega$ for the inductors, capacitors, and solder contacts respectively, we find that the experimental and simulation results match reasonably, as shown in Tables~\ref{table1} and \ref{table2}, and plotted in Fig.~\ref{fig2}d of the main text.

\subsubsection{Reducing the influence of parasitic resistances}
Parasitic resistance has a strong impact on the experiment. The most direct way to reduce its impact is to increase the inductances $L$ while decreasing the capacitances $C$, since doing so does not necessitate a proportional increase in the parasitic resistances. However, the inductance value should not be too large in order to keep R$_{pL}$ within a reasonable range. At the same time, if the capacitance value is too small, the equivalent series resistance of the capacitors becomes dominant and the frequency $f_0$ increases, which may increase the uncertainty in the measurement. In order to strike a balance, we chose \emph{C=4.7} nF, \emph{L=1} mH.

\subsection{Determination of $\omega_r$ for experimental setups}

Here, we provide details on how resistive contributions from $L$ and $C$ components (not necessarily parasitic) affect $\omega_r$, which is the most important dimensionless parameter in our setup. The addition of serial resistances to each capacitor and inductor modifies their admittance contributions to the circuit Laplacian as follows:
\begin{align}
i\omega C&\rightarrow \frac{i\omega C}{1+i\omega CR_C}\\
\frac{1}{i\omega L}&\rightarrow \frac{1}{R_L+i\omega L}
\end{align}

The Laplacian from the Eq.~1 of the main text is hence modified to
	\begin{widetext}
	\begin{align}
		\mathcal{L}(k_x,k_y)&=\frac{2i\omega C}{1+i\omega CR_C}(1-\cos k_x ) + \frac{2}{R_L+i\omega L}( 1-\cos k_y )\notag\\
		&=\frac{2i\omega C}{1+i\omega CR_C} \left[ (1-\cos k_x)-\frac{\frac{L}{C}-R_CR_L+i\omega LR_C + \frac{i R_L}{\omega C}}{R_L^2+\omega^2 L^2}(1-\cos k_y)\right]\notag\\
				&=\frac{2i\omega C}{1+i\omega CR_C} \left[ (1-\cos k_x)-\omega_r^{-2}(1-\cos k_y)\right]
	\end{align}
	\end{widetext}
with the important parameter $\omega_r^{-2}$ modified to
\begin{equation}
\omega_r^{-2}=\frac{\frac{L}{C}-R_C R_L}{R_L^2+\omega ^2 L^2}+\frac{\omega LR_C+\frac{R_L}{\omega C}}{R_L^2+\omega ^2 L^2}i.
\end{equation}
Substituting the measured parasitic resistances for our  fabricated circuits via $R_C=R_{pC}+2R_{pW}, R_L=R_{pL}+2R_{pW}$, and $\omega_r$ into the simulations, we find an excellent fit to the measured circuit impedances and their peaks (Fig.~\ref{fig2}d of the main text). Their corresponding $\omega_r$ are given in Table~\ref{table2}. Note that an imaginary part $\text{Im}\omega_r$ of $\sim 0.02$ to $\sim 0.06$ was acquired due to these resistances. Since the components used were not of particularly high quality, $\text{Im}(\omega_r)$ can potentially be reduced by one or more orders if necessary - in our case, they already suffice for demonstrating the anomalous impedance scaling.
\begin{table}
	\begin{tabular}{cccccllll}
		\cline{1-5}
		\multicolumn{1}{|c|}{\multirow{2}{*}{\begin{tabular}[c]{@{}c@{}}Z max at\\  N=\end{tabular}}} & \multicolumn{2}{c|}{sim.}                                  & \multicolumn{2}{c|}{exp.}                                  &  &  &  &  \\ \cline{2-5}
		\multicolumn{1}{|c|}{}                                                                        & $f_0(kHz)$                 & \multicolumn{1}{c|}{$\omega_r$}           & $f_0(kHz)$                 & \multicolumn{1}{c|}{$\omega_r$}           &  &  &  &  \\ \cline{1-5}
		\multicolumn{1}{|c|}{3}                                                                       & 127.0                  & \multicolumn{1}{c|}{$1.7297-0.0190i$} & 130.2                & \multicolumn{1}{c|}{$1.7733-0.0198i$} &  &  &  &  \\
		\multicolumn{1}{|c|}{4}                                                                       & 135.6                & \multicolumn{1}{c|}{$1.8468-0.0211i$} & 139.4                & \multicolumn{1}{c|}{$1.8986-0.0222i$} &  &  &  &  \\
		\multicolumn{1}{|c|}{5}                                                                       & 139.6                & \multicolumn{1}{c|}{$1.9013-0.0222i$} & 144.0                  & \multicolumn{1}{c|}{$1.9612-0.0234i$} &  &  &  &  \\
		\multicolumn{1}{|c|}{6}                                                                       & 243.8                & \multicolumn{1}{c|}{$3.3195-0.0600i$} & 250.4                & \multicolumn{1}{c|}{$3.4092-0.0630i$} &  &  &  &  \\
		\multicolumn{1}{|c|}{7}                                                                       & 256.3                & \multicolumn{1}{c|}{$3.4895-0.0658i$} & 263.8                & \multicolumn{1}{c|}{$3.9515-0.0695i$} &  &  &  &  \\ \cline{1-5}
		\multicolumn{1}{l}{}                                                                          & \multicolumn{1}{l}{} & \multicolumn{1}{l}{}                & \multicolumn{1}{l}{} & \multicolumn{1}{l}{}                &  &  &  & 
	\end{tabular}
	\caption{\textbf{The effective $\omega_r$ for various experimental data points and their simulated values (from Fig.~\ref{fig2}d of the main text)}. Due to parasitic resistances, $\omega_r$ acquires a small imaginary part on the order $10^{-2}$. }
	\label{table2}
\end{table}

\subsection{Data availability}
All data can be acquired from the corresponding authors upon a reasonable request.
\subsection{Code availability}
All code can be requested from the corresponding authors upon a reasonable request.

\section{References}
\bibliography{references}

\section{Acknowledgments}
	This work was supported by the Ministry of Education (MOE) of Singapore Tier-II (Grant Nos. MOE2018-T2-2-117 and MOET2EP50121-0014) and MOE Tier-I FRC (Grant Nos. A-0005110-01-00 and A-8000195-01-00). CHL acknowledges support from Singapore Ministry of Education's Tier I grant A-8000022-00-00. HS would like to thank the Agency for Science, Technology and Research (A*STAR) for its support of our research through the SINGA fellowship program.

\subsection{Author contributions}
C.H.L. initiated the idea, proposed the first draft of the manuscript, and supervised the entire project. H.S. conducted preliminary studies, performed theoretical calculations, and contributed to parts of the writing. Z.B.S. assisted in improving the theoretical analysis and formulations. X.Z. and B.Z. designed and executed the experiment, analyzed the data, and wrote the experimental details. S.M.R., J.F.K., B.S., R.T., and M.B.J. reviewed and contributed to all aspects of the manuscript.

\subsection{Competing interests}
The authors declare no competing interests.

\end{document}